# Diurnal variations of the count rates from dark photons in PHELEX.

A. Kopylov*, I. Orekhov, V. Petukhov

Institute for Nuclear Research, Russian Academy of Sciences,

117312 Moscow, Russia, Prospect of 60th Anniversary of October Revolution 7A

*Corresponding author

E-mail address: kopylov@inr.ru

ABSTRACT. This paper reports the PHELEX experiment, i.e., PHoton–ELectron EXperiment, to search for the dark photons (DPs) of cold dark matter using a multicathode counter technique specifically developed for this purpose by the authors. The paper presents new data, a novel upper limit for the constant of kinetic mixing, and the first results of measurements of the diurnal variations in solar and stellar frames. The perspectives of this method are outlined in terms of the search for DPs.

Keywords: dark matter, dark photons, diurnal variations

Undoubtedly, dark matter (DM) is one of nature's great enigmas, which poses a challenge for the present experiment. Solving the problem requires obtaining a DM signal and proving its DM origin. The importance of this issue was demonstrated by the results obtained by the DAMA/LIBRA experiment [1]; the annual variations of the observed signal in that research can be considered as proof that weakly interacting massive particle (WIMP)-nucleon scattering is observed in NaI(Tl) crystals. However, as independent experiments have failed to validate this result to date, this is still an open question. The task is both difficult and rewarding since it involves the evolution of the universe. Physicists are employing different techniques in an attempt to explore all possibilities in the search for DM, and numerous experiments have been performed to pursue this goal. The present paper describes the authors' work in the search for dark photons (DPs) as cold dark matter (CDM) using a novel technique proposed and developed by the authors specifically for this purpose. As the technique has been described previously [2, 3], this paper focuses on the search for diurnal variations of the count rate as a promising technique to prove that DPs are truly observed as CDM. The work has been done in the framework of PHoton–ELectron Experiment at the Institute for Nuclear Research, Moscow.

First, what is a dark photon? The discussion on the possible existence of a new Spin-1 Boson began last century in the early 1980s [4–6] and has resulted in extensive literature on the topic; review [7] can be referenced as a broad academic venture into this subject. From an experimental perspective, this object is an interesting focus of study as a new boson has a mass that can be transformed into the energy of a usual massless photon due to kinetic mixing with a dimensionless parameter ($\chi$), thereby quantifying the mixing. By this parameter, particles from the dark and standard model sectors are connected, which enables researchers to observe the effect. This hypothesis facilitates the introduction of a new field (A′) and two parameters ($\chi$ and $m_{\gamma'}$) into Lagrangian:

$$L = -\frac{1}{4} F_{\mu\nu} F^{\mu\nu} - \frac{1}{4} F'_{\mu\nu} F'^{\mu\nu} - \frac{\chi}{2} F_{\mu\nu} F'^{\mu\nu} + \frac{m_{\gamma'}^2}{2} A'_{\mu} A'^{\mu} \qquad (1)$$

where $A_\mu$ and $A'_\mu$ are the photon and DP fields, respectively; $F_{\mu\nu}$ and $F'_{\mu\nu}$ are the corresponding field strength tensors; and $m_{\gamma'}$ is the mass of a DP. With the introduction of two new parameters into the theory the effect of this can be observed in multiple ways. First, the most popular method of observing this particle is via its decay, e.g., to an e+e− pair [8] at accelerators where it can be created. Second, the effect of the scattering of the photon created from DPs in kinetic mixing can be observed on the valence electrons of atoms. In this case, the expected effect is proportional to the mass of the target, and the threshold and background count rates of the detector should be very low.

DAMA/LIBRA was one of the first experiments in this sphere and was succeeded by numerous experiments following the same ideology (e.g., reference [9] and others therein). Over the past three decades, these investigations, which were designed to search for WIMP scattering on the nuclei of a target, have made significant advances. Some of these detectors are also sensitive to the scattering of DPs on the valence electrons of atoms and notable experimental techniques have been employed, e.g., scintillating and semiconducting detectors, dual-phase xenon detectors, and cryogenic detectors. The upper limits for the scattering of DPs on the valence electrons of a target's atoms have been obtained (e.g., reference [10] and the references contained therein). The original method is an idea to use a dish antenna [11] to observe the reflected radiation from the surface of the antenna by the conversion of DPs into real ones. In this case, the effect is proportional to the surface of the antenna due to the power collected by the antenna, as follows:

$$P = 2\alpha^2 \chi^2 \rho_{CDM} A_{cath} \qquad (2)$$

where $\alpha^2 = \cos^2\theta$, $\theta$ is the angle between the vector of the E-field of the DP-originated photon and the surface of the antenna, $\rho_{CDM}$ is the energy density of DM (which can be taken here as $\rho_{CDM} = (0.55 \pm 0.12)$ GeV/cm$^3$ [12]), and A is the surface of the antenna. However, this method does not work at higher photon energies because as the coefficient of photon reflection from the surface of an antenna becomes low, the photon is absorbed rather than reflected. To apply this method to higher energies, the authors proposed the use of a proportional counter instead of an antenna to detect the electrons emitted in the photo effect from the surface of a counter's cathode by the conversion of DPs into real ones. In this case, the count rate of single electrons can be obtained using the following equation:

$$P = m_{\gamma'} R_{MCC} / \eta \qquad (3)$$

where $m_{\gamma'}$ is the mass of a DP, $R_{MCC}$ is the count rate, and $\eta$ is the quantum efficiency for a given cathode; here, this was taken to be equal to that of the real photon with energy equal to the mass of the DP. The upper limit is obtained by combining Eqs. (2) and (3).

$$\chi = 2.1 \cdot 10^{-12} \left( \frac{R_{MCC}}{\eta \cdot 1Hz} \right)^{\frac{1}{2}} \left( \frac{m_{\gamma'}}{1eV} \right)^{\frac{1}{2}} \left( \frac{0.55 GeV/cm^3}{\rho_{CDM}} \right)^{\frac{1}{2}} \left( \frac{1m^2}{A_{cath}} \right)^{\frac{1}{2}} \left( \frac{\sqrt{2/3}}{\alpha} \right) \qquad (4)$$

To subtract the background count rate from the events with traces of ionized particles cut at both ends of the counter, a multicathode counter was constructed [2, 3] (see Fig. 1 for its design).

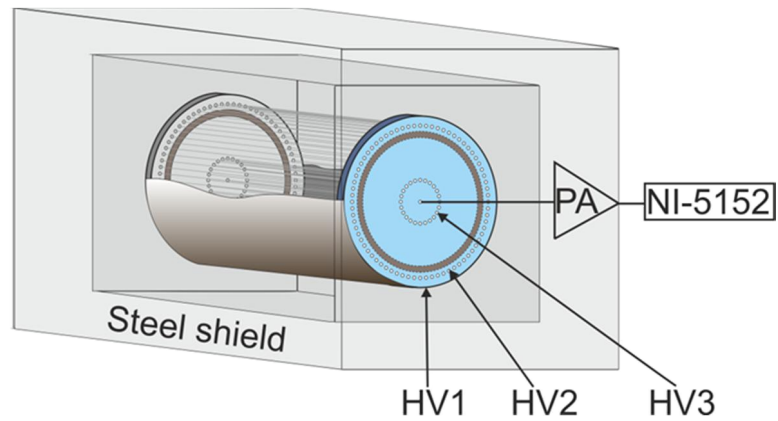

Figure 1. The scheme of multicathode counter. PA – preamplifier. NI-5152 – digitizing board. HV1-HV3 – potentials at the cathodes.

The counter has three cathodes instead of one. The first, which is an outer metallic cylinder, acts as a target. The second, which is fabricated from 50-μm-diameter wires located 5 mm from the first cathode, functions as a barrier to the electrons emitted from the first cathode. Finally, the third cathode, which is also made of 50-μm wires tightened in a 20-mm radius around a 20-μm-diameter anode wire, enables a high ($>10^5$) gas amplification; this is required to register the single electrons emitted from the outer cathode.

This study used two counting configurations. The first was utilized when the potential at the second cathode was higher than at the first. This configuration allows electrons emitted from the outer cathode to drift freely toward the anode. In addition to the electrons emitted from the outer cathode, the background from both ends of the counter was counted in this configuration.

In the second configuration, the potential applied to the second cathode was lower than that of the outer cathode; therefore, the electrons emitted from this cathode were rejected and unable to reach the counter's anode. Only the background was counted in this configuration. The net effect was determined by subtracting the count rate in the second configuration from that in the first configuration run by run. After many counting runs, the data were collected to obtain the result. This study's upper limit result of $\chi < 10^{-11}$ for the mass range from 9 to 40 eV obtained using a counter with an aluminum cathode and a Ne + $CH_4$ (10%) mixture published in [13] was included in the review of PDG in reference [14]. The sensitivity of this study's proposed method is highest within this mass range [3]. It should be noted that here, free electrons of a degenerate electron gas were used as a target; therefore, the proposed method complements those using valence electrons as targets. The details of the physics in the above processes differ, which may have significant implications in terms of data interpretation. Recently, measurements were made using the same counter with increased data. The measurements were made directly on the surface of Earth at the ground floor in a building in Troitsk, Moscow. In this case, the counter was crossed by multiple muons from cosmic rays. Figure 2 presents one of the frames taken with a pulse from a muon crossing the counter and a pulse from a single electron.

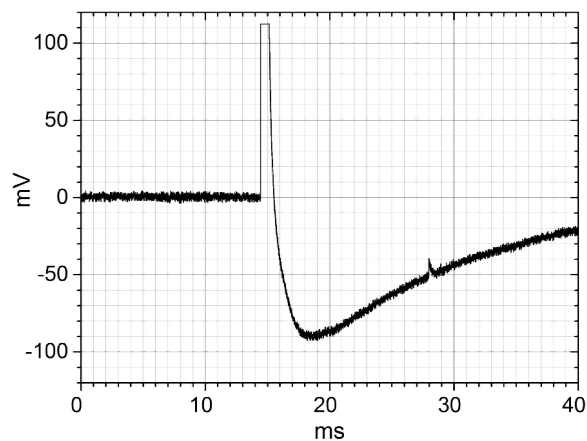

Figure 2. The pulses from muon (left) and from single electron (right).

The pulses from single electrons were collected at intervals between the pulses from muons. Because a typical count rate of muons was approximately 15 per second and a dead time after the leading edge of each muon pulse was approximately 10 ms, there was minimal counting interference. Around 1 TB of data was collected on a hard disk daily, and all data treatment was performed offline. Figure 3 shows the count rates measured in configurations 1 and 2.

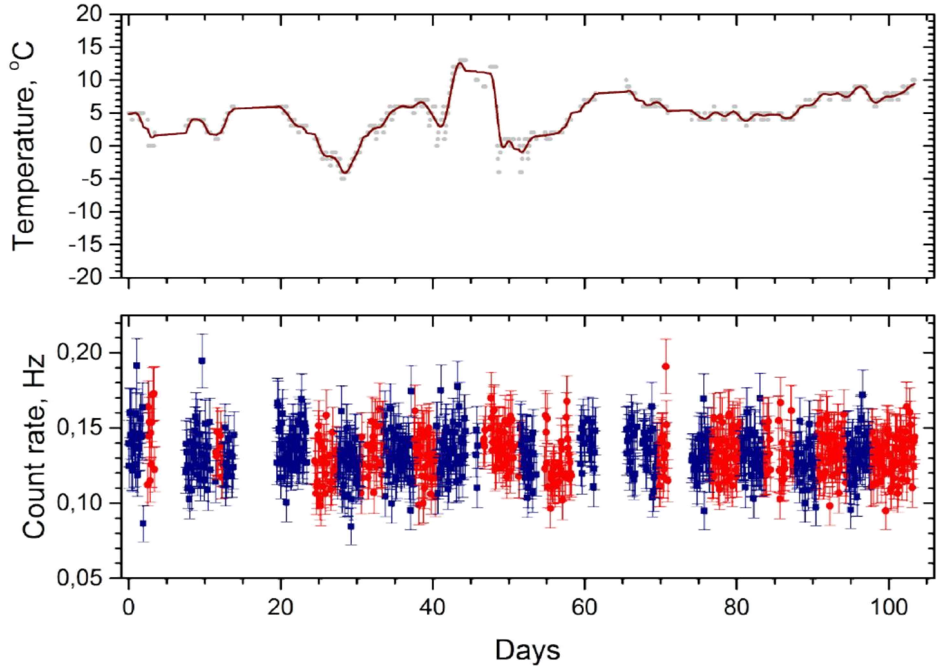

Figure 3. The count rates measured in configuration 1 (red) and configuration 2 (blue). Upper diagram shows the temperature of the counter during measurements.

The net count rate of the single electrons emitted from the outer cathode was obtained by subtracting the count rate in configuration 2 from that in configuration 1. Based on these measurements, the count rate from the single electrons was obtained, as follows:

$$r_{MCC} = -0.00018 \pm 0.00101 \text{ Hz}$$

and from here:

$$R_{MCC} = (r_{MCC} + 2\sigma)/\varepsilon = 0.0030 \text{ Hz}$$

where $\varepsilon = 0.608$ – efficiency of counting of single electron [3].

Then, from Eq. (4), an upper limit at 95% CL can be obtained:

$$\chi < 6 \cdot 10^{-12} \text{ for mass interval from 9 till 40 eV.}$$

The detector presented in this study has a remarkable feature of directionality. As Spin-1 DM candidates for DPs have an intrinsic polarization direction, the polarization of DP-originated photons and that of DPs should match. In Eq. (2), the factor $\alpha^2$ means that if the flux of photons is polarized, i.e., the vector of the E-field has a certain direction in a galactic or solar frame, then the Earth's rotation will cause variations in the effect of DPs. When the E-field vector is along the axis of the counter when $<\alpha^2> = 0$, this effect is minimal. However, when this vector is perpendicular to the axis of the counter, then $<\alpha^2> = 0.5$. Both the geographical latitude of the location site of the detector and the specific orientation of the counter influence the diurnal variations of the effects of DPs [15]. This can be observed if DPs are polarized and the counter's cathode has a mirrored surface. The internal surface of the cathode of this paper's counter can be seen in Fig. 4.

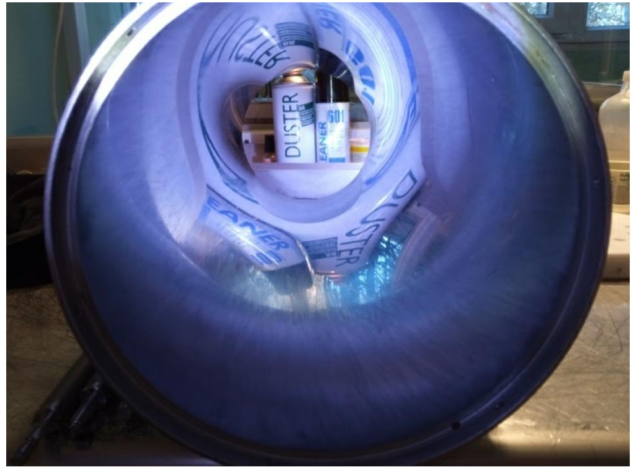

Figure 4. The internal surface of the cathode of multicathode counter.

For the control measurements, a counter with a matt cathode can be used to ensure that there is no effect in this case. The curve of the diurnal variations of the count rate can be acquired by collecting the results of the measurements in configurations 1 and 2 obtained at specific time intervals during one solar day (24 hours) or one sidereal day (23 hours, 56 minutes, and 4 seconds). Figure 5 presents the expected effect calculated for the Moscow-based counter (geographical latitude 55°45′N) for three different orientations of counters (vertical, horizontal west–east, and north–south) for different angles between the E-field vector and the Earth's axes. The starting-point moment was selected when the vector of the E-field was in the plane of the Moscow meridian.

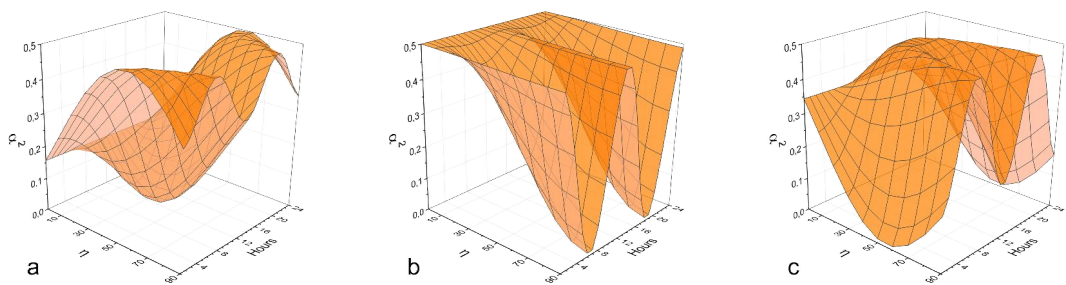

Figure 5. The expected diurnal variations for counters placed in Moscow for different orientations and different angle η between vector of E-field and the axes of the Earth.
(a) – vertical orientation, (b) east-west orientation, (c) north-south orientation.

Significant differences can be observed in the variation curves for different orientations and angles (η); however, they are always symmetrical relative to the moment of 12:00. This feature enables the suppression of curves without this symmetry as false curves. Additionally, it facilitates the identification of a moment in time when the vector of the field is in the plane of the corresponding meridian. Figure 6 shows the curves obtained for 871 points calculated from this study's latest measurements presented in Fig. 3.

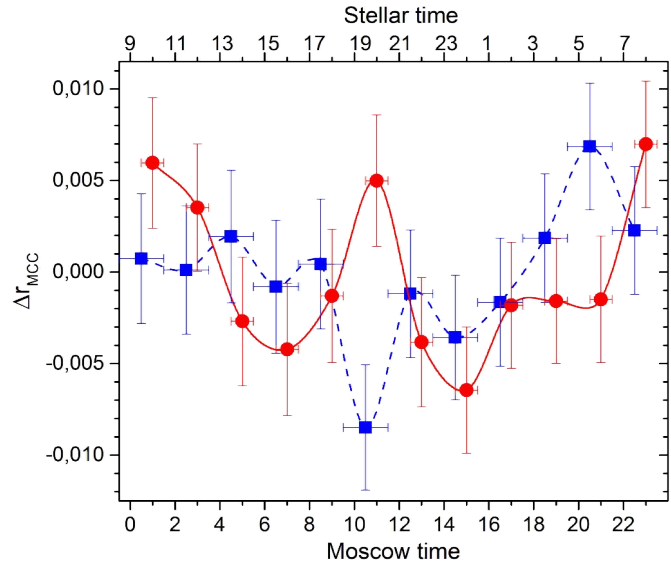

Figure 6. The variation curves obtained for solar day (red line and lower scale) and sidereal day (blue line and upper scale).

The error bars here are large, and no conclusions can be drawn; however, it illustrates what can be obtained from data analyses if the data are expanded. Accordingly, a moment of time relative to when the curve exhibits symmetry can be acquired. Suppose that, for the blue curve, this moment is at 18:00 stellar time. Then, it follows that, first, this study's results support the vector of the E-field having a certain direction within a stellar frame. Then, this moment of time relative to when symmetry is observed is 6 hours later the moment of 12:00. Second, this means that 6 hours later the moment of 00:00, i.e., at the moment of 06:00 stellar time, the E-field vector was in the plane of the Moscow meridian where the detector is situated at present. Then, the obtained and calculated curves should be compared for different angles between the vector of the E-field and the axes of the Earth for specific orientations of the measurement counter. If an agreement is found between the calculated and observed curves at some angles and specific orientations of the counter, then the orientation of the E-field vector can be obtained. Then, the factor $\alpha^2 = \cos^2\theta$ in Eq. (2) can be considered correct. However, if it is not, then something should be omitted from the considerations, or the factor differs from $\cos^2\theta$; consequently, a comparison should be made with the data obtained using counters at different orientations and potentially at different geographical latitudes. This could indicate the explanation.

This field encompasses an expansive study area in which the existence of symmetrical curves in observed diurnal variations is a critical point. Furthermore, for the control, counters with matt cathode surfaces should not exhibit this feature. These considerations will form the basis of the authors' future work.

**Declaration of competing interest**

The authors declare that they have no known competing financial interests or personal relationships that could have appeared to influence the work reported in this paper.

**Acknowledgements**

We appreciate very much the substantial support from the Ministry of Science and Higher Education of Russian Federation within the "Instrument Base Renewal Program" in the framework of the State project "Science".